\documentclass[12pt,preprint]{aastex}

\slugcomment{Article for Astrophysical Journal Letters}

\shorttitle{Planck-Kerr Turbulence}
\shortauthors{Gibson, C. H.}

\begin{document}


\title{Planck-Kerr Turbulence}


\author{Carl H. Gibson\altaffilmark{1}}
\affil{Departments of Mechanical and Aerospace Engineering and  Scripps
Institution of Oceanography, University of California,
     San Diego, CA 92093-0411}

\email{cgibson@ucsd.edu}


\altaffiltext{1}{Center for Astrophysics and Space Sciences, UCSD}


\begin{abstract} A quantum gravitational instability is identified at Planck
scales between  non-spinning extreme Schwarzschild black holes and spinning
extreme Kerr black holes, which produces a turbulent Planck particle gas.  
Planck inertial vortex forces balance gravitational forces as the Planck
turbulence cascades to larger scales and the universe expands and cools.
Turbulent mixing of  temperature fluctuations and viscous dissipation of
turbulent kinetic energy provide irreversibilities necessary to sustain the
process to the strong force freeze out temperature where inflation begins.
Turbulent temperature fluctuations are fossilized when they are stretched by
inflation beyond the horizon scale of causal connection.  As the horizon of the
expanding universe grows, the fluctuations seed patterns of nucleosynthesis, and
these seed the formation of structure in the plasma epoch.  Fossil big bang
turbulence is supported by extended self similarity coefficients \citep{bs02}
computed for cosmic microwave background temperature anisotropies that match
those for high Reynolds number turbulence.

\end{abstract}


\keywords{turbulence, cosmology}


\section{Introduction} Observations show that the statistical distribution of
galaxies and their clusters are chaotic and homogeneous, suggesting that
something about the big bang mechanism was capable of randomly producing and
randomly distributing the energy density seeds required to trigger
nucleosynthesis and gravitational structure formation in patterns
characteristic  of high Reynolds number turbulence.  Was the big bang
turbulent?  The big bang is presently the no-man's-land of modern physics. 
Quantum mechanics (QM) and general relativity (GR) theories develop
singularities under the big bang Planck scale conditions that have not yet been
reconciled by multidimensional superstring theory or any other theory of the
quantum gravitational dynamics (QGD) epoch.  The string tension in string theory
$c^{4}G^{-1} = 1.1 \times 10^{44}$ kg m s$^{-2}$ \citep{gr99} appears in
Einstein's equation and matches the Planck force
$F_P$ (see Table 1).

Turbulence theory is not in much better condition than QM and GR.  No definition
of turbulence is generally accepted, and it is widely believed that turbulence
always cascades from large scales to small, despite clear evidence to the
contrary in jets, wakes and boundary layers where turbulence length scales
monotonically increase.  How can turbulence exist at the time of the big bang
when there are no large scales of anything to supply the energy?  Furthermore,
observations of cosmic microwave background (CMB) temperature anisotropies
$\delta T/T \approx 10^{-5}$ prove the plasma before transition to gas was not
turbulent, since turbulence would produce much larger  $\delta T/T \approx
10^{-2}$ values
\citep{gib00}.  Whatever turbulence emerged from the big bang was damped by
processes in the plasma epoch.

In the following we apply the Planck-Kerr instability model to the QGD epoch
before inflation, using dimensional analysis.  It appears that not only did QGD
turbulence exist, but that turbulence was the key mechanism responsible for the
big bang.  Evidence of high Reynolds number turbulence preserved by the CMB is
discussed, and conclusions are presented.

\section{Theory} General relativity theory requires the length scale $L_{SS}$ of
black holes of mass
$m$ to approximate the Schwarzschild radius $L_{SS} \equiv Gm/c^2$, where $G =
6.67
\times 10^{-11}$ m$^3$ kg$^{-1}$ s$^{-2}$ is Newton's gravitational constant and
$c = 3 \times 10^{8}$  m$^{1}$ s$^{-1}$ is the speed of light.  Quantum
mechanical wave-particle duality assigns a de Broglie wavepacket size $L_{dB}
\equiv h/mc$ to particles of mass $m$, where $h \equiv 1.05 \times 10^{-34}$  kg
m$^{2}$ s$^{-1}$ is Planck's constant.  The GR and QM length scales match when
the two theories break down, where $L_{SS} \approx L_{dB}$ and the black hole
mass and particle mass equal the Planck mass $m_P \equiv [chG^{-1}]^{1/2} = 2.12
\times 10^{-8}$ kg.  Substituting $m_P$ gives the Planck length
$L_P \equiv [c^{-3}hG]^{1/2}  = 1.62
\times 10^{-35}$ m.  The Hawking evaporation time of a black hole and its de
Broglie wave propagation time
 $ c^{-1} L_P$ match at the Planck time $t_P \equiv [c^{-5}hG]^{1/2} = 5.41
\times 10^{-44}$ s. The Hawking temperature
$T_{H} \equiv [c^{3} h G^{-1}k^{-1}m^{-1}]$ of a Planck scale black hole (Planck
particle) is
$T_H = [c^{5}h G^{-1}k^{-2}]^{1/2} \equiv T_P$, where $T_P = 1.40
\times 10^{32}$ K is defined as the Planck temperature and $k = 1.38 \times
10^{-23}$  kg m$^{2}$ s$^{-2}$ K$^{-1}$ is the Boltzmann constant.  The energy
of Planck particles $E_P \equiv [c^5 h G^{-1}]^{1/2} = 1.94 \times 10^{9}$ kg
m$^{2}$ s$^{-2}$.  The Planck entropy $S_P \equiv  E_P / T_P = k$.

How could the quantum gravitational dynamics epoch be turbulent while we know
the plasma epoch at the CMB time of photon decoupling was not? Turbulence is an
eddy-like state of fluid motion where the inertial vortex forces of the eddies
are larger than any other forces that might tend to damp the eddies out
\citep{gib91}.  Large Planck inertial vortex forces per unit mass  
$[\vec{v}
\times
\vec{\omega}]_P \equiv [c^{7}h^{-1}G^{-1}]^{1/2} \equiv g_P = 5.7 \times
10^{51}$  m s$^{-2}$ must match and sometimes overwhelm the Planck gravitational
acceleration
$g_P$ as well as the Planck viscous forces for big bang turbulence to be
triggered and persist.  The size of the universe at the strong force freeze out
time
$t_{SF}
\approx 10^{-35}$ s is
$L_{SF}
\approx ct_{SF} = 3 \times 10^{-27}$ m at the strong force freeze out temperature
$T_{SF}
\approx 10^{28}$ K
\citep{pea00}.  The strong force velocity of the Planck particles $V_{SF}
\approx c[T_{SF}/T_{P}]^{1/2} = 3 \times 10^{6}$ m s$^{-1}$.  The Planck
viscosity $\nu_P = cL_P = [c^{-1}hG]^{1/2} = 4.8 \times 10^{-27}$ m$^{2}$
s$^{-1}$, assuming viscous stresses are transported by Planck  particles at
light speeds with Planck length scale collision distances.  Thus, Reynolds
numbers
$VL/\nu$ in the QGD epoch increase to
$ Re_{SF} \approx V_{SF} \times L_{SF}/\nu _{SF} \approx 10^6$ values that are
well above critical.  Small, weakly collisional particles like gluons, neutrinos
and photons that inhibit turbulence by viscous stresses after the strong force
freeze out cannot exist at QGD temperatures.

The Planck-Kerr mechanism for big bang turbulence formation is shown in Figure
1.  Vacuum oscillations at Planck length scales allow a small possiblity for
Planck particles and Planck anti-particles to appear spontaneously by quantum
tunneling.  Once they appear they increase the possibility of more Planck
particles to be triggered by their enormous temperatures.  Stable spinning
quantum states analogous to positronium are possible (extreme Kerr black
holes).  Prograde accretions of Planck particles on these objects may trigger a
turbulent Planck gas with vorticity matching the Plank-Kerr spin, taking
advantage of the increased prograde energy release available from small
marginally stable prograde orbits (Peacock 2000 p61) to increase the probability
of further Planck-Kerr and Planck particle formations. Energies up to 42\% $E_P$
are available.

The Planck-Kerr instability of Figure 1 has Planck power $P_P \equiv c^5 G^{-1}
= 3.64 \times 10^{52}$ kg m$^2$ s$^{-2}$, which is $10^{4}$ larger than the
power of all galaxies in our horizon at the present time.  The Planck turbulence
stress per unit mass
$[\vec{v}
\times
\vec{\omega}]_P / {L_P}^2 = \sigma _ P 
\equiv [c^{13} h^{-3} G^{-3}]^{1/2} = 1.26 \times 10^{121}$ m$^{-1}$ s$^{-2}$. 
From general relativity theory, space-time is created by sufficiently large
negative stresses.  Planck-Kerr turbulence supplies the power, stresses and
entropy required to initiate space-time creation and reach the strong force
phase transition state where negative stresses of the false vacuum can trigger
inflation
\citep{gu97}.

\section{Observations}

Fingerprints of high Reynolds number turbulence have been observed in the CMB
temperature anisotropies using extended self similarity (ESS) coefficients
\citep{bs02}.  Comparisons with high Reynolds number turbulence ESS coefficients
and with ESS coefficients of other flows \citep{ben96} are shown in Figure 2
\citep{gib01}.  The CMB and turbulence ESS coefficients are in nearly perfect
agreement with each other, but do not match ESS coefficients computed from
non-turbulent flows.  

\section{Conclusions}

A variety of indicators suggest that the big bang was not only turbulent but
strongly turbulent before inflation, even though that time period was short and
the length scales small.  No strong sources of irreversibility besides
turbulence and turbulent mixing exist at Planck scales where the temperatures
are  high and highly reversible vacuum oscillations produce little or no
entropy. The Planck specific entropy $s_P \equiv [c^{-1}h^{-1}G k^{2}]^{1/2}$
was only $6.35 \times 10^{-16}$ m$^{2}$ s$^{-2}$ K$^{-1}$.  The turbulent
viscous dissipation rate  $\varepsilon _P \equiv [c^9 h^{-1} G^{-1}]^{1/2}$ was
$1.72 \times 10^{60}$ m$^{2}$ s$^{-3}$.

The Planck-Kerr instability is identified between extreme Schwarzschild black
holes and extreme Kerr black holes that can produce small scale vorticity and an
expanding universe driven by  Planck inertial vortex forces and turbulence
stresses that overcome Planck gravity, stetch space, and create more turbulent
Planck gas.  Strong big bang turbulence mixes the resulting temperature
fluctuations as the universe cools to the strong force phase transition
temperature.  Inflation fossilizes the turbulent temperature fluctuations by
stretching the length scales beyond the scale of causal connection.  

Fossil big bang temperature turbulence fluctuations retain their turbulence
signatures by seeding nucleosynthesis, and are detected by observations of the
homogenous chaotic pattern of large scale cosmic structures, and the extended
self similarity coefficients and dissipation multiscaling parameters of the
cosmic microwave background temperature anisotropies that identically match
those of high Reynolds number turbulence and mixing.

\acknowledgments

\clearpage

\begin{figure}
        \epsscale{0.8}
        \plotone{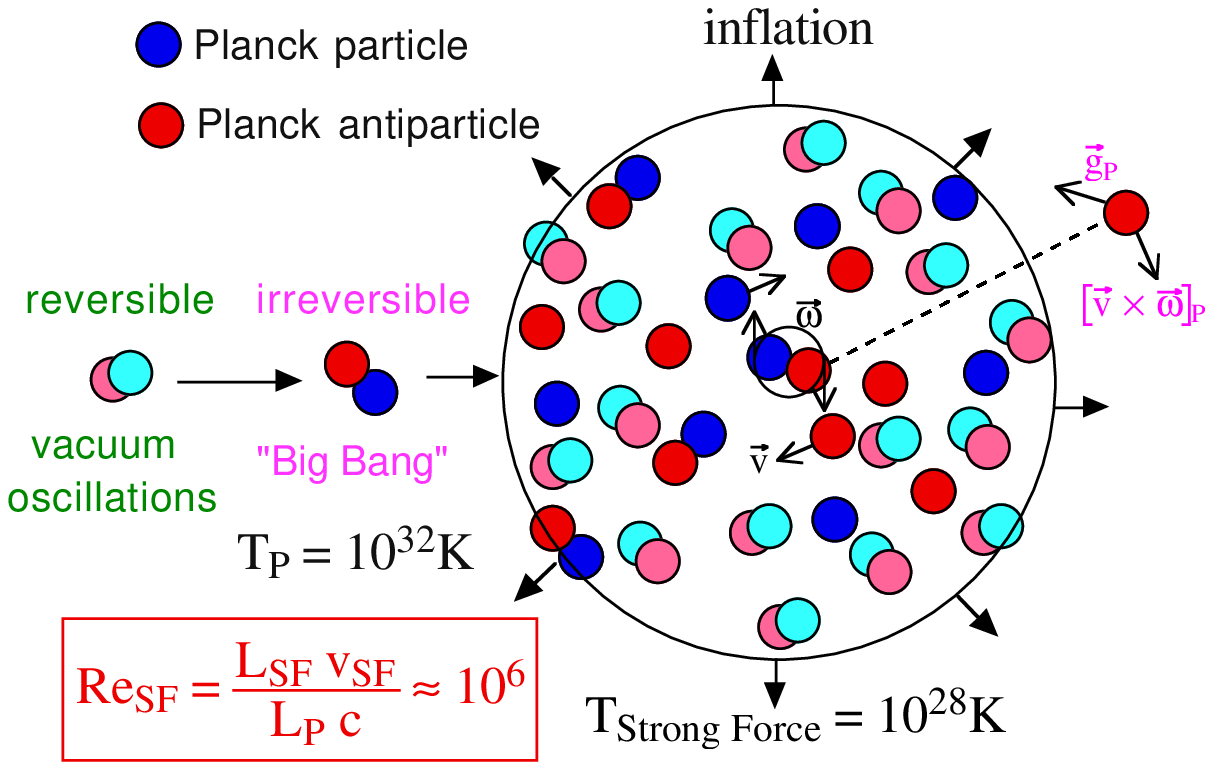}
        %%\plottwo{epsfile}{epsfile}
        \caption{Planck-Kerr mechanism  for big bang turbulence \citep{gib01}. 
Vacuum oscillations at Planck scales produce a Planck particle and a Planck
anti-particle by quantum tunneling, which trigger more such particles by the
same mechanism.  If a spinning Planck-Kerr particle (an extreme Kerr black hole)
forms, the possibility exists to produce a rotational turbulent gas of Planck
particles from the large energy released by prograde accretions 
\citep{pea00}.  When the expanding space-time cools to $T_{SF} = 10^{28}$ K (the
strong force freeze out temperature), the high Reynolds number turbulent
temperature fluctuations are fossilized by inflation stretching their scales
beyond the horizon scale of causal connection
$c t_{SFo} \approx 3 \times 10^{-25}$ m at the end of inflation $t_{SFo} \approx
10^{-33}$ s \citep{gu97}.}
       \end{figure}

\begin{figure}
        \epsscale{0.6}
        \plotone{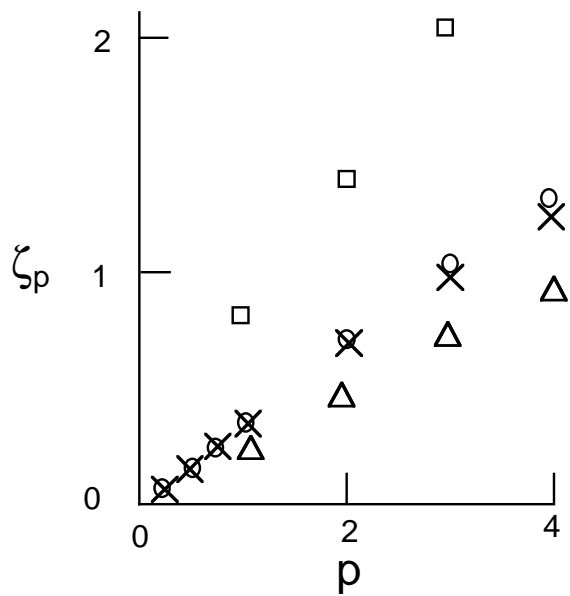}
        %%\plottwo{epsfile}{epsfile}
        \caption{Extended Self Similarity coefficients $\zeta _p$ (ESS)  for CMB
temperature anisotropy $\delta T$ differences $\delta T (r)$,  where $r$ is the
separation distance for the p$^{th}$ order temperature structure functions 
$\langle \vert \delta T  (r) \vert ^p \rangle \sim r^{-\zeta _p}$ (circles,
Bershadskii and Sreenivasan 2002) compared to high Reynolds number turbulence
($\times$'s), the solar wind (triangles), and Reyleigh-Benard convection
(squares) from Benzi et al. 1996, Table 1.}
       \end{figure}

\clearpage

\begin{deluxetable}{lrrrrcrrrrr}
\tablewidth{0pt}
\tablecaption{Planck-Kerr Turbulence Scales and Parameters}
\tablehead{
\colhead{Scales and Parameters}& \colhead{Symbol}           &
\colhead{Definition$^a$}      &
\colhead{Value}           }
\startdata
\sidehead{Scale} 
Planck mass & $m_P$ &$[chG^{-1}]^{1/2} $& $ 2.12
\times 10^{-8}$ kg
\\
Planck length & $L_{P}$&$[c^{-3}hG]^{1/2} $& $ 1.62
\times 10^{-35}$ m
\\  Planck time & $t_{P}$&$[c^{-5}hG]^{1/2} $& $ 5.41
\times 10^{-44}$ s
 \\ Planck temperature & $T_{P}$&$[c^{5}hG^{-1}k^{-2}]^{1/2} $& $ 1.40
\times 10^{35}$ K
\\
Planck entropy & $S_{P}$&$k$& $ 1.38 \times
10^{-23}$  kg m$^{2}$ s$^{-2}$ K$^{-1}$
\\
\sidehead{Parameter}
Planck energy & $E_P$ &$[c^{5}hG^{-1}]^{1/2} $& $ 1.94
\times 10^{9}$ kg m$^{2}$ s$^{-2}$
\\
Planck power & $P_P$ &$c^{5}G^{-1} $& $ 3.64
\times 10^{52}$ kg m$^{2}$ s$^{-3}$
\\
Planck dissipation rate & $\varepsilon_P$ &$[c^{9}h^{-1}G^{-1}]^{1/2} $& $ 1.72
\times 10^{60}$  m$^{2}$ s$^{-3}$
\\
Planck density & $\rho_P$ &$c^{5}h^{-1}G^{-1} $& $ 5.4
\times 10^{96}$ kg m$^{-3}$ 
\\
Planck specific entropy & $s_P$ &$c^{-1}h^{-1}Gk^{2} $& $ 6.35
\times 10^{-16}$  m$^{2} s^{-2} K^{-1}$ 
\\
Planck force & $F_P$ &$c^{4}G^{-1} $& $ 1.1
\times 10^{44}$ kg m s$^{-2}$
\\
Planck gravity force & $g_P$ &$c^{7}h^{-1}G^{-1} $& $ 5.7
\times 10^{51}$  m s$^{-2}$
\\
Planck inertial-vortex force & ${[\vec{v}
\times
\vec{\omega}]_P}$ &$c^{7}h^{-1}G^{-1} $& $ 5.7
\times 10^{51}$  m s$^{-2}$
\\
Planck viscosity & $\nu_P$ &$[c^{-1}hG]^{1/2} $& $ 4.8
\times 10^{-27}$  m$^{2}$ s$^{-1}$
\\
Planck Reynolds stress & $\sigma_P$ &$[c^{13}h^{-3}G^{-3}]^{1/2} $& $ 1.3
\times 10^{121}$  m$^{-1}$ s$^{-2}$
\\

%\cutinhead{This is a cut-in head}
%\sidehead{I am a side head:}

\enddata
\tablenotetext{a}{$c = 2.99 \times 10^{8}$  m s$^{-1}$ is the speed of
light, $h
= \hbar
\equiv 1.05 \times 10^{-34}$  kg m$^{2}$ s$^{-1}$ is Planck's constant,
$G = 6.67
\times 10^{-11}$ m$^3$ kg$^{-1}$ s$^{-2}$ is Newton's gravitational constant, $k = 1.38 \times
10^{-23}$  kg m$^{2}$ s$^{-2}$ K$^{-1}$ is the Boltzmann constant.}

%% You can append references to a table using the \tablerefs command.

%\tablerefs{}
\end{deluxetable}

\end{document}